\documentclass[aps,prl,superscriptaddress,twocolumn,floatfix,nofootinbib]{revtex4}
\usepackage{amssymb}
\usepackage{amsmath}
\usepackage{epsfig}
\usepackage{dcolumn}
\usepackage{rotating}
\usepackage{color}
\usepackage{hyperref}
\definecolor{rosso}{cmyk}{0,1,1,0.4}
\definecolor{rossos}{cmyk}{0,1,1,0.55}
\definecolor{rossoc}{cmyk}{0,1,1,0.2}
\definecolor{blu}{cmyk}{1,1,0,0.3}
\definecolor{blus}{cmyk}{1,1,0,0.6}
\definecolor{bluc}{cmyk}{1,1,0,0.1}
\definecolor{verde}{cmyk}{0.92,0,0.59,0.25}
\definecolor{verdec}{cmyk}{0.92,0,0.59,0.15}
\definecolor{verdes}{cmyk}{0.92,0,0.59,0.7}

\newcommand{\ba}{\begin{eqnarray}}
\newcommand{\ea}{\end{eqnarray}}
\newcommand{\be}{\begin{equation}}
\newcommand{\ee}{\end{equation}}
\newcommand{\bi}{\begin{itemize}}
\newcommand{\ei}{\end{itemize}}

\newcommand{\la}{\lambda}

\newcommand{\en}{\epsilon}

\newcommand{\La}{\Lambda}

\newcommand{\cF}{{\cal F}}
\newcommand{\cP}{{\cal P}}

\newcommand{\ra}{\rightarrow}
\newcommand{\Ra}{\Rightarrow}

\newcommand{\LF}{\left(}
\newcommand{\RF}{\right)}

\newcommand{\Rd}{\right.}

\newcommand{\mx}{\mbox}
\newcommand{\mt}{\mathtt}

\newcommand{\for}{\mx{ for }}

\newcommand{\QG}{\mt{q}}
\newcommand{\sech}{\mt{sech}}

\begin{document}

\title{Super-Inflation, Non-Singular Bounce, and Low Multipoles  }

\author{Tirthabir Biswas}
\affiliation{Physics Department, Loyola University, Campus Box 92, New Orleans, LA 70118}
\author{Anupam Mazumdar}
\affiliation{Consortium for fundamental physics, Lancaster University, Lancaster, LA1 4YB, United Kingdom}


\begin{abstract}
In this paper we present a simple argument that shows a non-singular bouncing cosmology naturally yields an era of
super-inflation which can precede the phase of normal potential driven inflation. One of the consequences of a super-inflation
phase is that it might be able to account for suppressing the low multipole in the amplitude of the cosmic microwave background
radiation. We are able to constrain the number of e-folds of super-inflation from the current Planck data to be roughly 3 e-folds in a model independent  way.
\end{abstract}

\maketitle


The recent results from WMAP~\cite{WMAP}  and Planck~\cite{Planck:2013kta} on cosmic microwave background (CMB) radiation
suggest lack of power at large angular scales or at very low multipoles, i.e. $l\leq 40$. Although these data points are well within our cosmic variance and statistically their significance is still low, the power deficit is not insignificant,  around $5-10\%$ at $2.5-3~\sigma$ as compared to the 6 parameter fit of $\Lambda {\rm CDM}$ model~\cite{Planck:2013kta}. This lack of amplitude of power spectrum on very large angular scales  is  intriguing and perhaps it might be shedding some light on yet to be understood physics at the earliest epochs.

Since the temperature anisotropy is typically given by the fluctuations in the scalar field whose energy density is dominating the early universe. In the
simplest scenario of a slow roll inflation, %
\begin{equation}\label{exp-1}
{\cal P}^{1/2}_\Phi\sim\frac{\delta T}{T}\sim \frac{\delta \rho}{\rho}\sim 10^{-5}\,,
\end{equation}
Naively,  one might expect to increase $\rho$ while keeping the fluctuations $\delta \rho$ nearly constant in order 
to suppress the power spectrum. However this procedure fails. Typically, increasing $\rho$ during inflation would be a daunting task.
In an absence of any other degrees of freedom, the energy density of the inflaton cannot in general increase unless we are willing to 
violate the coveted energy conditions~\footnote{Some exceptions have been noted. In principle on can try to play with both $\delta\rho$
and $\rho$ to make the power spectrum blue shifted and then red shifted. This can be arranged in multi field scenarios, such as ~\cite{assisted},
or in stochastic inflation~\cite{Riotto}, where some Hubble patches increase their energy density while some lower, or via fast roll inflation which 
precedes the slow-roll inflation~\cite{Linde-Peloso}.  The latter scenario suppresses the power spectrum during the fast roll phase by keeping 
the Hubble parameter constant while increasing the velocity of the scalar field which suppresses the amplitude, see~\cite{fast-roll}.
The {\it inflection point} inflation models~\cite{visible-1} can also lead to modify the spectral tilt at low multipoles from blue-to-red tilt as shown in Ref.~\cite{jain}. 
}.

The aim of this paper is to provide a very simple mechanism for suppressing the low multipoles by an era of {\it super-inflaion}, with $d H(t)/dt >0$.
The concept of super-inflation contradicts the common intuition. One cannot increase the value of $H(t)$ as it would mean an increase in
$\rho(t)$ in an expanding universe.  A rapid acceleration can only dilute the energy density or the number density, even a cosmological constant is only able to retain it's constancy of vacuum energy density. The challenge is to pump even more energy  into the universe, but from where?

Such a scenario cannot be achieved within Einstein's theory of general relativity (GR). One needs to modify GR in the ultraviolet (UV) in order to yield a phase of super-inflation. At low energies, or in the infrared (IR), then one would naturally obtain
GR with all its success, which will also keep a subsequent phase of inflation within the realm of GR.

One advantage of super-inflation will be that it can modify the spectral tilt at low $l$ by making it towards {\it blue-tilt} before the conventional slow
roll inflation kicks in with a decreasing $H(t)$, which is known to yield a {\it red-tilted} spectrum. The prediction for a super-inflating phase arises in any non-singular bouncing cosmology which we will discuss below.

Any mechanism which attempts to explain the cosmological singularity problem within a semi-classical  space-time description
will naturally yield an era of {\it super inflation} where $H(t)$ is increasing with time. By semi-classical we simply mean that we 
are able to approximately formulate the dynamics of {\it quantum gravity} in terms of a metric theory of space-time, possibly with 
higher derivative corrections. If such a description remains valid at energy scales we are interested in, then there are only two 
ways to avoid the singularity to prevent the scale factor of the universe approaching $0$, i.e. $a(t)\ra 0$~\footnote{In Ref.~\cite{copeland} it was 
argued that in loop-quantum gravity one can also yield super-inflation. }:

\begin{itemize}
\item{Non-singular bounce: where we can have a bounce where the universe transits from a contracting to an expanding phase at some finite cosmological time.}

\item{Emergent universe: we can have an emergent universe~\cite{emergent} where as $t\ra-\infty$, $a(t)\ra$ non-zero constant. }

\end{itemize}

In either case there exists a phase when $\dot{H}>0$. For a bouncing case this can be seen very easily -- $H(t)$ is precisely zero at the bounce,
and then it becomes more and more positive after the bounce:
\begin{equation}\label{at bounce}
\dot a =0\,,~\ddot a>0\Ra \dot{H}={\ddot{a}\over a}-{\dot{a}^2\over a^2}>0\,~~{\rm~at~the~bounce}\,,
\end{equation}
 where $'a'$ is the scale factor and dot denotes time derivative w.r,t the physical time, $'t'$.  
 
Since a non-singular bounce arises from the modification of GR, hence this trend in increasing $H(t)$  continues till the modifications of GR fades away gradually in the IR ($\dot{a},\ddot{a},\dddot{a}\dots$ becoming small compared to the scale of quantum gravity, which could be presumably one or two orders of magnitude smaller than the Planck scale), thus leaving only GR in IR. Once this happens there is a turn-over in the expansion rate of the universe,  $H(t)$ starts to decrease, entering the usual slow-roll  phase of inflation due to the slowly rolling inflaton, for a review see~\cite{infl-rev}, which must be embedded within the visible sector~\cite{visible,visible-1,visible1} in order to match the success of thermal history of the universe. 

We are aware of one particular non-singular bouncing scenario that can be constructed analytically within a string-theory inspired nonlocal higher derivative extension
 of Einstein's GR which is {\it covariant} and {\it ghost free} construction, see~\cite{BMS,BKM,BKMV,Koshelev:2012qn,alex}, which we will briefly mention below.
 
In order to understand intuitively how a super-inflationary phase can explain the deficit in the primordial power spectrum seen by both WMAP~\cite{WMAP} and Planck~\cite{Planck:2013kta}, let us first consider perturbations  of a  canonical scalar field which obeys the Klein-Gordon equation:
\begin{equation}
\ddot{\delta\phi_k}+3H\dot{\delta\phi_k}+\left[\left(\frac{k}{a}\right)^2+V''(\phi)\right]\delta\phi_k=0\,,
\end{equation}
Here 'dot' denotes derivative w.r.t. time and 'prime' denotes derivative w.r.t. $\phi$. We emphasize that this equation does not rely on GR and  therefore remains valid even if we modify GR in the UV. It is well known that approximately the amplitude of any given {\it comoving} mode freezes once it crosses the Hubble radius at a value given by
\begin{equation}
{\delta \phi}_{k} \propto {H\over k^{3/2}}
\label{delta-phi}
\end{equation}
provided that these perturbations were seeded by the quantum Bunch-Davis vacuum fluctuations of the scalar field  deep in the UV/sub-Hubble regime. The above 
Eq.~(\ref{delta-phi}) leads to a slightly {\it red-tilted}  power spectrum, ${\cal P}_{\phi}$, which is given by:
\begin{equation}
{\cal P}_{\phi}^{1/2}\propto {H_k\over M_p}\ ,
\label{power-spectrum}
\end{equation}
 in the standard slow-roll inflationary models. The {\it red tilt} $(n_s<1)$ arises from the fact that the Hubble parameter slowly decreases with time as the scalar field gently rolls downwards to its potential, and $H\propto \sqrt{V}$. $H_k$, refers to the Hubble parameter when a given mode crosses the Hubble radius. This yields a well-known form of the power-spectrum:
\begin{equation}
{\cal P}_{\phi}=A\LF{k\over k_p}\RF^{n_s-1}
\end{equation}
where $k_p$ denotes the pivot scale, and the spectral tilt $n_s$ is given in terms of the slow roll parameters:
\begin{equation}\label{tilt}
n_s=1+2\eta -6\epsilon\,,~ \epsilon\equiv \left({M_p V'\over V}\right)^2\,,~~ \eta\equiv{M_p^2V''\over V}\,.
\end{equation}

From Eqs.~(\ref{delta-phi},~\ref{tilt}), it is clear that if we have a phase of super-inflation where $H(t)$  increases with time for a certain period,  the tilt of the spectrum would be mostly blue $(n_s>1)$. Let us consider a simple illustrative example with a scale factor
which is {\it geodesically complete}:
\be
a(t)=\cosh \la t\ ,
\label{a}
\ee
which is a solution arises within nonlocal theories of gravity, first pointed out in Ref.~\cite{BMS}. 
For Eq.~(\ref{a}) the Hubble parameter becomes:
\be
H(t)=\la\tanh\la t
\ee
So, for large times the space-time asymptotes to a de-Sitter universe, $a(t)\sim e^{\la t}$, and $H$ approaches $\la$ from {\it below.}, increasing with time.
To reiterate, one would definitely need to modify GR in the ultraviolet (UV) in order to realize such a solution, but as soon as
GR becomes the dominant theory in the infrared (IR), the Hubble parameter would reach a maximum value
and then it will start to decrease according to the GR dynamics.

Now we wish to discuss the transition from a blue ($n_s>1$) to a red-tilted ($n_s<1$) spectrum expected in any bouncing scenarios.
To simplify our discussion let us first note that in most modified gravity scenarios, one can capture the effects of the ``quantum gravity terms'' as additional contributions in the energy momentum tensor (see, for instance Ref.~\cite{alex}):
\be
\dot{H}=-{1\over 2M_p^2}(\rho_{\phi}+p_{\phi}+  \rho_{\QG}+p_{\QG})\,,
\ee
where we have now included the  higher derivative terms of the scale factors contained in the modified Einstein tensor as an effective energy density, $\rho_{\QG}$, and its corresponding pressure, $p_{\QG}$, along with  the usual energy density and pressure contributions from the inflaton field $\phi$.
The modification in GR in the UV, $\rho_{\QG},~p_{\QG}$, is mainly responsible for driving a super-inflationary phase where $\dot{H}>0$.

For a hyperbolic cosine bounce such as Eq.~(\ref{a}), this can be directly computed from the behaviour of the bouncing scale factor:
\be
|\rho_{\QG}+p_{\QG}|\approx 2 M_p^2\la^2\sech^2\la t
\ee
One expects a transition from the UV regime to slow-roll inflationary GR regime to occur once the driving force from the inflation field, controlled by
\be
|\rho_{\phi}+p_{\phi}|=\dot{\phi}^2\ ,
\ee
overwhelms the UV regime:
\be
\dot{\phi}^2\approx 2 M_p^2\la^2\sech^2\la t
\ee
Now, one can express the left hand side in terms of the slow roll parameter and the potential energy, $V\approx 3M_p^2\la^2$:
\be
\dot{\phi}^2=\LF{V'\over 3H}\RF^2=\LF{V'^2M_p^2\over 3V}\RF^2={2\over 3}\en V=2\en M_p^2\la^2\ ,
\ee
so that the epoch of transition is given by an implicit relation~\footnote{According to our convention, $k$, has a physical meaning, it is the physical wave number for a given mode at the bounce point.}
\be
\en=\sech^2\la t={1\over 1+\LF{k_{\max}/ \la}\RF^2}\,.
\ee
As we can see during the super-inflation phase $H(t)$ increases, which would indicate an increase in the power as larger $k$ modes exit the Hubble patch till $k=k_{\max}$, after which we would get the usual red-tilted power-law spectrum. 

There is however one subtle point in going over from  ${\cal P}_\phi$ given by Eq.~(\ref{power-spectrum}) to the power spectrum for metric fluctuations 
$\cP_{\Phi}$, where $\Phi$ being the metric potential. In the standard inflationary cosmology  one has to use GR equations for this purpose. This relation can become modified if one goes beyond the GR paradigm, which one must in order to obtain a bounce. However,
as already argued, the modes which are of phenomenological interest, exit the Hubble radius just before the commencement of the GR regime and therefore is expected to have minimal modifications.

Moreover, in many bounce models, such as based on non local  modifications to GR~\cite{BMS,BKM}, the perturbation equations remains unchanged for the regimes in question~\cite{BKMV}, as a consequence of the fact that these theories do not introduce any new degrees of freedom~\cite{BKMV}. With the above caveat in mind we will in this paper continue to use Eq.~(\ref{power-spectrum}), a more detailed study would need the specifics of the different bounce mechanisms in question, which we leave for future investigation.

Let us now estimate the expected spectrum for the hyperbolic cosine bounce. Although for definiteness we use Eq.~(\ref{a}), the same algorithm can be used to obtain the spectrum for any other type of bounce. In order to obtain the spectral form of $\cP_{\Phi}$, one needs to express $H(t)$ as a function of $k$. This can be done by using the Hubble crossing condition:
\be
k=aH\Ra k=\la \sinh\la t
\label{hubble-crossing}
\ee
Using the above, the power spectrum can be written as
\be
\cP_{\Phi}\propto {{k/ \la}\over \sqrt{1+\LF{k/\la}\RF^2}}
\ee
Matching the super-inflationary power spectrum with the inflationary power spectrum at $k=k_{\max}$ we have the following form:
\be\label{cp}
\cP_{\Phi}=\left\{
\begin{array}{lr}
A\LF{k\over k_{\max}}\RF\sqrt{{1+\LF{k_{\max}/ \la}\RF^2\over 1+\LF{k/ \la}\RF^2}}& \for k<k_{\max}\\
A\LF{k\over k_{\max}}\RF^{\eta_s-1}& \for k>k_{\max}
\end{array}
\Rd
\ee
So, we have introduced two new parameters, $k_{\max}$ and $k_{\max}/\la$.

Let us make an estimate to see whether such a power spectrum can explain the low multipole moments. The observed reduction of power, around 10\%, is over $l=2$ to $l=40$ range, so that approximately over $k_{\max}/k\sim 10$. Using the power-spectrum Eq.~(\ref{cp}), we have the constraint
\be
0.9\approx \LF{1\over 10}\RF\sqrt{{1+\LF{k_{\max}\over \la}\RF^2\over 1+{1\over 100}\LF{k_{\max}\over \la}\RF^2}}\,.
\ee
There indeed exists a solution and the corresponding value of the slow roll parameter:
\be
\LF{k_{\max}\over \la}\RF^2\sim 400\Ra \en \sim 2.5\times 10^{-3}\ ,
\ee
This is completely consistent with an inflationary cosmology. What the above numbers also tell us is that the transition happens  in the near 
de-Sitter phase away from the bounce. If $t_0$ is the transition time, then from Eq.~(\ref{hubble-crossing}), we find the number of e-foldings,
${\cal N}$, during the super-inflation
\be
e^{\cal N}\sim \sinh \la t_0\approx 20\Ra a_0\sim a_0=\cosh\la t_0 \approx 20
\ee
In other words, by  the time the $l\sim 40$ mode exits, the universe has grown by a factor of $20$ ( ${\cal N}\sim 3$ e-folds) from it's minimal bounce size. 
One can also calculate $\la$ from the observed amplitude of spectrum:
\be
10^{-5}= {\la\over \sqrt{\en} M_p}\Ra \la \sim 10^{-6}M_p\sim 10^{12}~{\rm GeV}\,.
\ee   
This scale will be relevant for building any visible sector model of inflation which will be a case for future investigation.

To complete the story, let us briefly discuss a particular  higher derivative gravitational theory that is known to produce a hyperbolic cosine bounce considered here. Any resolution to the cosmological singularity problem must give rise to a {\it covariant} and {\it ghost free} theory of gravity in the UV, a general criteria for constructing such theories was obtained in~\cite{BGKM}. These typically contain an infinite series of higher derivative terms (and hence the association with nonlocality) often in the form of an exponential, which are rather ubiquitous in string theory~\cite{witten}.
Both cosmological background solutions and their perturbations  have been studied in great details for a particular sub-class of these theories  in 
Refs.~\cite{BMS,BKM,BKMV}, whose actions have the form
\begin{equation}
 S=\int d^4x\sqrt{-g}\left(\frac
 {M_P^2}{2}R+R{\cal F}(\Box/M_*^2)R-\Lambda+ \mathcal{L}_\mathrm{M}\right),
 \label{nlg_action}
\end{equation}
where $\Lambda$  is the cosmological constant which is rather straight forwardly related to the $\la$ parameter: $\la=\sqrt{\La/3M_p^2}$. $M_{\ast}$ is the mass scale at which the higher derivative
terms in the action become important and typically $M_*\sim \La^{1/4}$. 
$\mathcal{L}_\mathrm{M}$ is the matter Lagrangian. $\cF$ is an analytic function:
\begin{equation}
{\cal F}(\Box/M_*^2)=\sum\limits_{n\geqslant0}f_n\Box^n\,,\\
\end{equation}
where:  $\Box \equiv g^{\mu\nu}\nabla_\mu\nabla_\nu$=$g^{\mu\nu}\nabla_\mu\partial_\nu$= $\frac{1}{\sqrt{-g}} \partial_\mu \left(\sqrt{-g} \, g^{\mu
  \nu}\partial_\nu \right)$,
and $\nabla_\mu$ is the covariant derivative.
For energy scale below the UV cut-off $M_{\ast}$, the higher derivative theory of gravity asymptotes to the standard GR, and all the higher derivative terms become irrelevant at the infra red (IR) limit. What is particularly relevant for us is that in Ref.~\cite{BKMV} it was shown that the perturbations behaved exactly like the metric fluctuations in GR as space-time approached the deSitter limit, which is where most of our relevant modes will be exiting the Hubble-radius. Another intriguing feature of this model was that we found that the ``effective'' stress energy tensor coming from the higher derivative terms have a non-vanishing anisotropic stress at the perturbative level.

To summarise, we have argued that an era of super-inflation preceding normal inflation can possibly render the blue spectrum ($n_s >1$) for the cosmological perturbations which might help to explain the low multipoles of the CMB observed in WMAP~\cite{WMAP} and Planck~\cite{Planck:2013kta}. However the concept of
super-inflation cannot be realized in the context of GR, one requires an UV  modification of GR, which leads to a non-singular bouncing cosmology. Therefore super-inflation is a generic prediction, or,  a signature of a non-singular bouncing cosmology, which can help us to understand the very early epochs of the universe just at the time when scalar potential driven inflation has kicked in.

{\it Acknowledgements:} TBÕs research has been supported by the LEQSF(2011-13)-RD-A21 grant from the Louisiana Board of Regents.
AM is supported by the Lancaster-Manchester-Sheffield Consortium for Fundamental Physics under STFC grant ST/J000418/1.


\end{document}